# Broadband dielectric studies in linseed oil:
# Supercriticality and the New Insight into Melting/Freezing Discontinuous Transition

Aleksandra Drozd-Rzoska, Sylwester J. Rzoska*, Joanna Łoś

[1]Institute of High Pressure Physics Polish Academy of Sciences, ul. Sokołowska 29/37, 01-142 Warsaw, Poland

(*) Corresponding author emails: sylwester.rzoska@unipress.waw.pl

**Abstract**

The long-range supercritical changes of dielectric constant, resembling ones observed in rod-like liquid crystalline compounds, are evidenced for the linseed oil on appraching liquid – solid discontinuous phase transition. The 'supercriticality' can be an additional factor supporting unique pro-health properties of linseed oil. It can be also important for numerous technological applications.

The report also reveals properties significant for the grand melting/ freezing discontinuous phase transition cognitive puzzle. Broadband dielectric spectroscopy revealed strong and long-range premelting (heating from the solid phase) and post-freezing (cooling from the liquid phase) effects, with critical-like parameterizations. They can be correlated with the 'grain model' for premelting effect, and its developments by Lipovsky model. The evidence for post-freezing effect, with fair critical-like portrayal, may indicate a specific granular solidification, associated with pretransitional fluctuations. Numerous hallmarks of liquid-liquid and solid-solid transition were also found. Notably, the melting temperature surrounding is related to the minimum and freezing temperature to the maximum of dielectric loss factor $D = tan\delta$. Regarding dynamics, three primary relaxation times have been found. Their relative changes in subsequent temperature intervals, related to the apparent activation enthalpy, follow critical-like patterns with the same singular temperature. For relaxation times evolutions, it leads to optimal parameterizations via the *'critical&activated'* equation, recently proposed.

**Introduction**

The extraordinary pro-health properties of flax seeds and linseed oil have been known since ancient times. Hippocrates, the 'father' of medicine (~ 400 BCE), advised it as a significant food remedy. Near ~ 800 CE, King & Emperor Charles the Great (Charlemagne) indicated it as substantial for a healthy life in an edict (!). In modern times, Mahatma Gandhi recommended linseed oil, and it is still strongly advised by doctors and nutritionists [1-3]. Nowadays, the pro-health properties of linseed oil are explained by specific features of their components [1-5]. This report shows that yet another, so far overlooked, factor may be important: supercriticality.

Supercriticality means the presence of exceptional properties in the broad surrounding of the critical point, i.e., a continuous phase transition [6-10]. They result from the appearance and dominance of multi-molecular fluctuations associated with approaching the new phase. Their lifetime $\tau_{fl.}$, and size (correlation length: $\xi$) show the singular, critical behavior [9-11]:

$$\xi(T) = \xi_0 |T - T_C|^{-\nu} \quad (1a) \qquad\qquad \tau_{fl.}(T) = \tau_{fl.}^0 |T - T_C|^{z\nu} \quad (1b)$$

where $T_C$ is the critical temperature, $\nu$ is the critical exponent for the correlation length, and $z$ is the so-called dynamic exponent: $z = 2$ for the conserved and $z = 3$ for the non-conserved order parameter. The 'critical quasi-phase' emerging near the critical point led to innovative supercritical technologies, strongly developing due to their 'green' features [6-12]. They are related to the selective extraction of componenta or selective chemical reactions, which strength can be tuned by the distance from $T_C$. The supercriticality is indicated as a promising technology for innovative food or pharmaceutical preservation or processing [6-8]. Supercriticality can also promote the adhesion of some dissolved agents to the surface of solid microelements, leading to their encapsulation [12].

For supercritical technologies, essential are two model-relations. The first one is the Kirkwood equation linking chemical reaction rate $k$ and solubility $s$ to dielectric constant $\varepsilon$ [13, 14]:

$$k, s = p_\infty \exp\left[\frac{A\Delta_P}{RT}\left(\frac{1}{\varepsilon} - 1\right)\right] \quad (2)$$

where $\Delta_P$ is for the difference in polarity between the reactant and product; $p_\infty$ is the prefactor related to the given property, $A$ is the system-dependent constant.

The second one, is the Noyes-Whitney dependence [13, 14]:

$$\frac{dm}{dt} = S\frac{D}{d}(C_S - C_B) \quad (3)$$

where $m$ is the mass of dissolved material, $t$ – is the processing time, $S$ is the surface area of the solute particle, $D$ diffusion coefficient, $d$- is the thickness of the concentration gradient layer, $C_S$ and $C_B$ are for dissolving particle surface and bulk concentrations (*mol/L*).

In ref. [10] the authors (SJR, ADR) converted the latter to the form containing the DC electric conductivity $\sigma$, relatively simple for experimental determining:

$$\frac{dm}{dt} = \left(\frac{k_B}{nq^3}\frac{S}{d}\right)T\sigma(T)(C_S - C_B) = KT\sigma(T)(C_S - C_B) \quad (4)$$

where $n$ is for the number of electric carriers/charges ($q$), $K = const$, $k_B$ is the Boltzmann constant and the coefficient $K = const$.

The above brief resume indicates that supercriticality, could be an important factor supporting, if not shaping, unique properties of linseed. However, neither communications nor even suggestions regarding supercritical phenomena in linseed oil have appeared so far.

This report reveals the weakly-discontinuous nature of linseed oil melting/freezing discontinuous transition, preceded by 'supercritical' long-range pretransitional changes of dielectric constant, including the low-frequency extension.

Broadband dielectric spectroscopy [15] studies also revealed the glassy- dynamics [16] with a unique universal portrayal. For the solid phase, the premelting effect (heating from the solid phase) [17, 18] and exceptional post-freezing effect (cooling from the liquid phase) have been found. They show well portrayed critical-like temperature changes. Consequently, the results presented can be also important for the melting/freezing discontinuous transition cognitive challenge [17-21].

## 2. Methods and Materials

Broadband dielectric spectroscopy (BDS) [15, 16] studies were carried out using Novocontrol Alpha impedance analyzer supplemented with Novocontrol Quattro temperature control unit. The complex dielectric permittivity $\varepsilon^* = \varepsilon' - i\varepsilon'$ scans in the frequency range $1 Hz < f < 10 MHz$, with 5-6 digits resolution, and 230 tested temperatures for ca. 200 K temperature range, were carried out. The real component of dielectric permittivity was determined as $\varepsilon'(f) = C(f)/C_0$, where $C$ is the capacitance for the measurement capacitor filled with the tested dielectric and $C_0$ is for the empty capacitor. The imaginary part was calculated as $\varepsilon''(f) = 1/2\pi f R(f) C_0$, where $R(f)$ is for the frequency-related resistivity [15]. Figure A1 in the Appendix shows obtained dielectric permittivity spectra, indicating characteristic regions, manifested in frequency scans. For the static domain (in liquids, usually $1 kHz < f < 1 MHz$), the frequency shift does not lead to a significant change of $\varepsilon'(f)$ [15]. Above the static domain there is the relaxation domain. Its hallmark is the primary relaxation $\varepsilon''(f)$ loss curve which peak (maximum) estimates the primary relaxation time for basic orientational processes: $\tau = 1/2\pi f_{peak} = 1/\omega_{peak}$. It can be estimated using the derivative of experimental data and the condition: $d\log_{10}\varepsilon''(f = f_{max})/d\log_{10}f = 0$ [16]. A strong rise of dielectric permittivity occurs for frequencies below the static domain. It can be used for determining the relative electric conductivity: $\sigma'(f) = \omega\varepsilon''(f) = 2\pi f \varepsilon''(f)$ [15]. The appearance of the horizontal, frequency-independent behavior in $\log_{10}\sigma'(f)$ vs. $\log_{10}f$ in the low-frequency (LF) domain determines the DC electric conductivity $\sigma'(f) = \sigma$ [15, 21]. Such behavior is presented in Figures A2 and A3 in the Appnedix. Distortions from DC-electric conductivity-related behavior can be linked, for instance, to the polarization of electrodes via Maxwell – Wegner effect [15], directly coupled to the translation of 'free' ionic species/contaminations. Notable, that DC electric conductivity is coupled to the primary relaxation time via the Debye-Stokes-Einstein (DSE) law $\sigma(T)\tau(T) = const$. For complex systems, with some 'local structure' the fractional fDSE law appears: $\sigma(T)[\tau(T)]^S = const$, where $S \neq 1$ denotes the fractional (decoupling) exponent [21, 22].

The strong rise of the real part of dielectric permittivity $\varepsilon'(f)$ in the LF domain is often explained heuristically as the impact of non-defined residual ionic contaminations [15, 21]. In the opinion of the authors, the phenomenon can be also explained by translational motions of basic molecules, which supports the coupling between translational and orientational motion given in DSE and fDSE laws [22].

BDS studies were carried out using the flat-parallel capacitor with gold-covered plates made from Invar: diameter $2r = 20mm$, and the distance between plates $d = 0.4\ mm$. The basic voltage applied in BDS measurements: $U = 1V$. The setup (Quattro Novocontrol unit) enabled the temperature control $\Delta T = \pm 0.02K$. Studies were carried out for frequencies in the middle of the static domain for subsequent temperatures: in practice, shifting from ~10 kHz to ~1kHz on cooling from 360 K to 120 K (see the Appendix). It allowed for avoiding biasing impacts of the low-frequency (LF) effects and high-frequency (HF) relaxation processes located below and above the static domain.

Preliminary DSC scan tests of the linseed oil between ~220 K and 450 K were carried out using the homemade apparatus, using $m \sim 1g$ of sample and copper-constantan thermocouples. The results are presented in Figure 1, overlapping those reported earlier [23-26]. Figure 2 DSC includes higher temperatures $T > 450\ K$, based on ref. [24], to complete the picture. This domain is beyond the topic of the report. For the given report, the most important is the strong manifestation of the melting/freezing temperature.

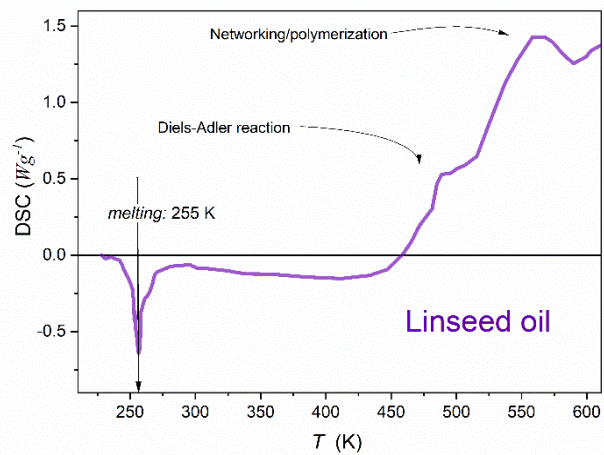

Fig. 1 **Differential Scanning Calorimetry (DSC), heat capacity related, scan in linseed oil**. Characteristic features significant for applications are indicated.

Regarding the tested material, golden flax seeds from the widely cultivated in Poland blue-flowered flax species *Linum usitatissimum*, were used. The linseed oil was obtained via cold-pressing of seeds in the laboratory. Seeds were purchased from Herbapol S.A., Lublin, Poland: the leading Polish company specializing in high-quality pro-health seeds & herbs products. The triple filtering of the oils using coffee filters was carried out.

Nowadays, it is well proved that linseed oil is an amazing pro-health food additive and a skin conditioner for natural cosmetics [27-33]. It can help wound healing and shows anti-inflammatory and regenerative properties [33, 34]. It supports heart health by acting towards maintaining the required cholesterol level [29]. There are suggestions for its stimulation impact on the immune system. The positive influence of linseed oil is broadly recognized in the fight against obesity, arthritis, and hypertension [1,4, 32, 34]. Let's focus on cosmetics-related issues: it can strengthen skin, support redness reductions, repair skin damage, including flaky epidermis, heal skin burns and frostbite, and create smooth and soft skin [1, 4, 32-34]. Linseed is also indicated as a prophylactic anti-cancer agent for breast, colon, and prostate problems [32-34].

These extraordinary properties are explained by the impact of unique linseed/flaxseed oil components and their bio-activity. Thet are polyunsaturated fatty acids (PUFA), phytoestrogenic lignans (secoisolariciresinol diglucoside, SDG), and an array of antioxidants such as phenolic acids and flavonoids. The beneficial PUFA of flax lipids are α-linolenic acid (ALA), (30–70% of the total fatty acid content), linoleic, (20% of the total fatty acid content), and oleic acid (30% of the complete fatty acid content) [3, 5, 32]. So far, the mentioned pro-health properties of linseed oil are explained by the beneficial properties of the mentioned components, supported by their synergic interactions possible due to excellent natural proportions [1-4, 32-34].

But this is only a part of linseed oil applications importance. It shows a polymer-forming tendency, especially at higher temperatures or in the presence of appropriate activators [35]. Therefore, it is used as an impregnator and varnish component for wood [36] and is a significant component of oil paints and putty hardening supports [37]. Already in the 19[th] century, it was used to produce linoleum, which is still one of the most important methods of covering floors [38]. In recent decades, Linseed oil has become the crucial component for vegetable oil-based electric insulation oils in electric power transformers: eco-friendly and able to maintain preferable properties even above 10 years [39].

## 2. Results and Discussion
### 2a. The static domain: supercriticality and the solid state premelting and post-freezing effects

There is broad evidence of dielectric studies [40-46] in edible vegetable oils, including linseed oil. They mainly focused on heuristic assessments of the material quality and were carried out in limited frequency and temperature ranges. There are neither evidence nor suggestions for pretransitional or supercritical phenomena in such systems, including linseed oil.

Figure 2 presents evidence for strong and long-range supercritical changes of dielectric constant in liquid linseed oil on cooling. Generally, the $solid \leftarrow liquid$ phase transition reached on cooling from the liquid phase is named the freezing temperature $T_f$, and its value depends on the cooling rate. It links $T_f$ to supercooling. The 'constant' material characterization is the melting temperature $T_m > T_f$ reached on heating from the solid phase, i.e., for $solid \rightarrow liquid$. However, for linseed oil, the same value of $T_f$ were reached for any applied cooling rate. It facilitated reliable isothermal frequency-related BDS scans. It also indicates that for linseed oil, one can consider two material characteristics (melting temperatures) associated with $solid \leftarrow liquid$ phase transition: $T_m^{cool} = T_f$, and $T_m^{heat} = T_m$ for $solid \rightarrow liquid$ scans. A similar behavior was observed, for instance, in liquid crystalline isopentylcyanobiphenyl (5*CB) [47, 48].

The key feature of the pretransitional/supercritical anomaly of dielectric constant is the crossover $d\varepsilon/dT > 0 \leftarrow d\varepsilon(T_{cross})/dT = 0 \leftarrow d\varepsilon/dT < 0$ on cooling, as visible in Figs. 2 and 3. The evolution of dielectric constant resembles the pretransitional, supercritical, anomaly in the isotropic liquid (I) phase of rod-like liquid crystalline (LC) compounds on approaching LC mesophases (M), such as the Nematic (N), Chiral Nematic (N*), Smectic A (SmA), Smectic E (SmE) [47, 49-52]:

$$\varepsilon(T) = \varepsilon^* + a(T - T^*) + A(T - T^*)^\phi \qquad (5)$$

where $T > T^* = T_{I-M} - \Delta T^*$, $T^*$ denotes the extrapolated temperature of a hypothetical continuous phase transition and $\Delta T^*$ is the metric of the phase transition discontinuity: it can range from $\Delta T^* = 1 - 2K$ for the I-N transition [49] to $\Delta T^* \sim 30K$ for I-SmE transition [51]; the exponent $\phi = 1 - \alpha$, where $\alpha$ describes the pretransitional anomaly of the heat capacity; in the given case $\alpha = 1/2$ and it is related to the mean-field or mean-field tricritical phase transitions.

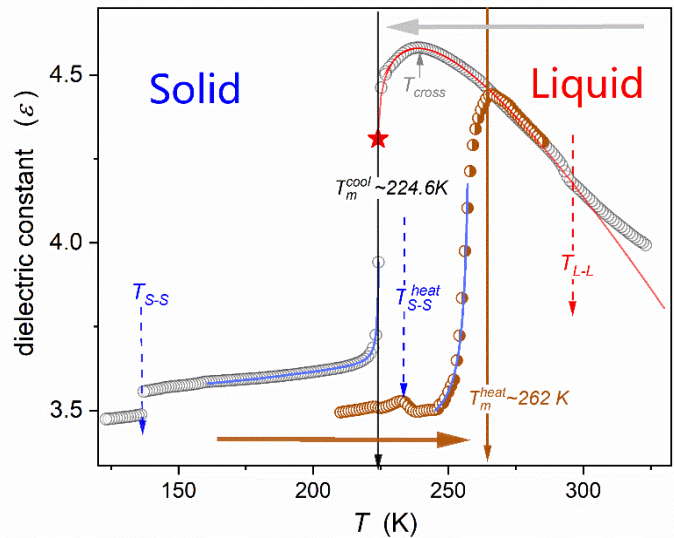

Figure 2  **Temperature changes of dielectric constant on cooling and heating in linseed oil.** Note the explicit manifestations of pretransitional effects on both sides of the freezing and melting temperatures. The results are for cooling (blue, open circles) and heating (orange, semi-filled circles). The portrayal of pretransitional effects is related to Eqs. (5) and (7), respectively. Solid vertical arrows indicate freezing and melting temperatures, respectively. The blue indicates the location of the continuous phase transition extrapolated from the liquid phase. The dashed vertical arrow indicates changes in the form of the temperature evolutions, which can eventually be associated with liquid–liquid or solid–solid transitions. The emergent solid-solid (S-S) and liquid-liquid (L-L) transitions are also indicated.

In LC materials Eq. (5) is associated with the prenematic arrangement of permanent dipole moments within pretransitional fluctuations, appearing due to the weakly discontinuous character of the phase transition. For rod-like molecules with a permanent dipole moment approximately parallel to the long molecular axis prenematic arrangement leads to the statistical antiparallel ordering of dipole moments. Consequently, their impact on the value of the dielectric constant within fluctuations is cancelled. The following relation between dielectric constant within fluctuation and the isotropic liquid surrounding takes place: $\varepsilon_{fluct.} \ll \varepsilon_{surrounding}$. The critical rise of the correlation length of fluctuations leads to the rise of their volume:

$$V_{fl.} \propto [\xi(T)]^3 \propto (T - T^*)^{-3\nu} = (T - T^*)^{-3/2} \qquad (6)$$

where $\nu = 1/2$ (mean-field value).

On approaching $T_{I-M}$ discontinuous phase transition, and hence also $T^*$, the volume occupied by fluctuations increases, finally leading to its prevalence and the behavior described by $d\varepsilon/dT < 0$. The temperature of the isotropic liquid – LC mesophase discontinuous transition can be recognized as the symmetry-limited melting temperature, i.e., $T_{I-M} = T_m$.

Eq. (5), with the same critical exponent $\alpha = 1/2$ fairly describes pretransitional (supercritical) behavior of dielectric constant in linseed oil, up to at 80 K above the liquid-solid transition. It is shown in Figure 2. The weakly discontinuous phase transition can be associated with a singular temperature, linked to a hypothetical continuous phase transition: $T^* = T_m^{cool} - \Delta T^*$. Notable is a little value of the discontinuity metric: $\Delta T^* \approx 0.6$.

**Table I** Values of parameters describing pretransitional changes of dielectric constant (Fig. 3 and Eq. (5) and its derivative (Fig. 4 and Eq. 6) for liquid linseed oil.

| Parameter | $\varepsilon^*$ | $a$ | $A$ | $\phi$ | $T^*$ | $T_m^{cool}$ | $\Delta T^*$ |
|---|---|---|---|---|---|---|---|
| Value | 4.310 | -16.4 | 125 | 0.50 | 224 | 224.6 | 0.6 |

The portrayal of experimental data presented via Eq. (5) was supported by the derivative-based analysis, for which the following description of the supercritical (in the liquid phase) anomaly can be expected:

$$\frac{d\varepsilon(T)}{dT} = a + (1 - \alpha)A(T - T^*)^{-\alpha} \qquad (7)$$

Such analysis reduces the number of fitting parameters and it is distortions-sensitive, leading to a strong manifestation of pretransitional effects, as shown in Figure 3. The curves describing the supercritical effects in Figures 2 and 3 are based on the simultaneous data analysis using Eqs. (5) and (6), in Figs. 2 and 3. The values obtained are given in Table I.

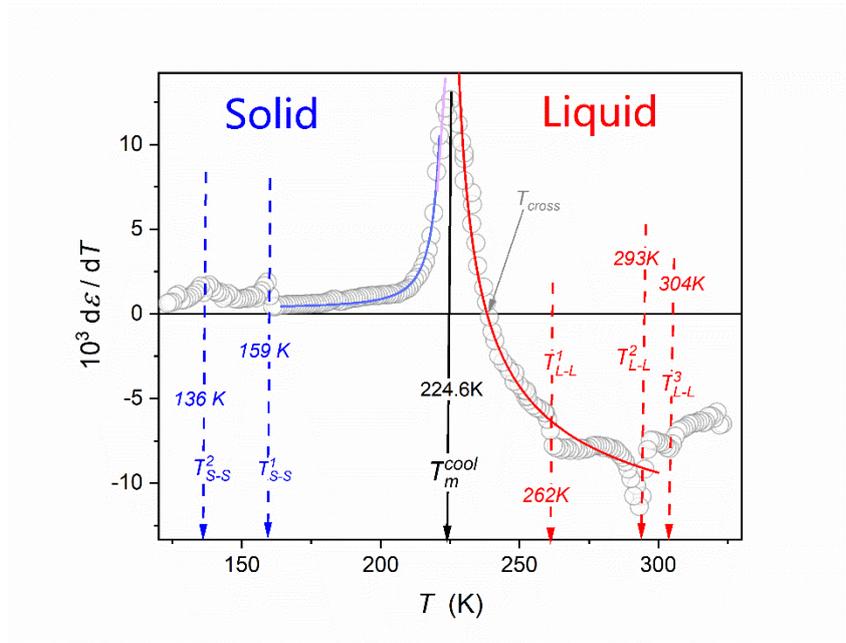

Figure 3  **The temperature evolution of the derivative of dielectric constant, detected on cooling in linseed oil.** Curves portraying pretransitional and premelting effects are related to Eqs. (6) and (8). Note the extraordinary manifestation of the sequence of solid-solid (S-S) and liquid-liquid (L-L) transitions.

The manifestation of the discussed pretransitional effects is very weak for $\varepsilon(T)$ changes registered on heating from the solid phase. It is associated with the fact that in the given case the liquid phase is available for $T > T_m = T_m^{heat}$, located well above the crossover temperature $T_{cross}$. Notable, that a similar behavior occurs for the mentioned case of liquid crystalline 5*CB [47, 48].

Generally, no pretransitional effects are expected in the solid phase for melting or freezing discontinuous phase transitions [17]. Notwithstanding, weak and range-limited premelting effects (i.e., on heating from the solid phase) are often observed. The related evidence can be concluded as follows [17, 21, 53-62]: (i) premelting effects are 'weak' and range-limited, (ii) they are detected for some systems, (iii) they are detected for some physical properties, (iv) there are no reports on their parameterization, (v) no solid-state pretransitional effects for freezing (cooling from the liquid phase). Despite the apparent limitations and problems, premelting effects are commonly considered the gate for explaining the nature of melting/freezing discontinuous phase transitions [17, 53-62]. This cognitive situation is qualitatively different from continuous phase transition: understanding their origins is one of the greatest universalistic successes of the 20th-century physics, concluded in the *Critical Phenomena Physics*. Eqs. (1), (5), and (6) are based on this concept. Essential for the success were the long-range and well-characterized pretransitional effects [11].

For melting/freezing discontinuous transition, such pretransitional effects are (almost) absent, except for the puzzling premelting effects in the solid state [17, 21, 53-63]. In linseed oil, both premelting and post-freezing effects exist, and the latter is even stronger and long-ranged. This is evident in Fig. 3, for the derivative transformation of experimental data.

The following relation well portrays the premelting and post-freezing effects in linseed oil:

$$\chi(T) = \varepsilon(T) - 1 = a + bT + \chi^0(T_C^* - T)^{-1} \quad , \text{ on heating for } \quad T < T_m \tag{7}$$

where $\chi$ denotes dielectric susceptibility, directly coupled to the polarization of the dielectric, the prefactor $\chi^0 = const$, and $T_C^*$ is the 'critical-like' temperature almost coinciding with $T_m$; the background term $c = const$.

For the derivative of dielectric constant:

$$\frac{d\varepsilon(T)}{dT} \approx \chi_0(T_C^* - T)^{-2} \quad , \quad \text{on cooling for } T < T_f = T_m^{cool} \tag{8}$$

The simultaneous analysis of experimental data presented in Figs. 2 and 3 yielded parameters collected in Table II and shown by solid curves.

**Tab. II** Values of parameters related to the post-freezing effect in the solid state, described by Eqs. (7),(8), with the discontinuity metric $\Delta T^{**} = T^{**} - T_m^{cool}$.

| Parameter | $a$ (K$^{-1}$) | $b$ | $M$ (K$^{-1}$) | $\phi$ | $T^{**}$(K) | $T_m^{cool}$ (K) | $\Delta T^{**}$ (K) |
|---|---|---|---|---|---|---|---|
| Value | $9.3 \times 10^{-4}$ | 3.83 | 0.159 | -1.0 | 224.75 | 224.6 | 0.15 |

Very recently, the behavior described by Eqs. (7) and (8) were reported by the authors (ADR, SJR) for premelting effect in nitrobenzene [63]. For explaining the results, firsthe grain model introduced for the premelting effect and discontinuous transition was recalled [57-62]. It links the premelting effect to grains appearing in the solid-state on heating towards $T_m$. Their appearance is linked to rising vibrations and defects within crystalline materials, finally leading to fragmentation and melting. Quasi-liquid nano-layers cover solid/crystalline grains. Lipovsky [64-66] developed the model focused on liquid nano-layers and showed their critical-like properties, with the 'critical' temperature $T_m^*$ almost coinciding with $T_m$. He derived a set of critical-like relations describing the pretransitional evolution of liquid nano-layers. For the order parameter-related susceptibility:

$$\chi_T(T) = \chi_0(T_m^* - T)^{-\gamma} \tag{9}$$

where the exponent $\gamma' = \gamma = 1$, if the mean-field approximation obeys in the given system.

Unfortunately, the liquid nano-layers occupy a 'very tiny' volume of the material in a premelting domain, in comparison with solid grains. It yields the essential experimental problem for the detection focused on liquid nano-layers, with the resolution enabling temperature evolution scans. However, in ref. [63] it was noticed that the BDS response from the liquid dielectric associated with the dielectric constant is qualitatively more significant than from the solid/crystalline state, which enables the 'extraction' of the contribution from liquid nano-layers for a sample in the premelting region.

For liquids, nano-constraints can create a quasi-negative pressure, i.e., stretching ('rarefication'), which can weaken the dipole-dipole short-range interactions for dipolar liquid. It may lead to the creation of conditions approapriate for the description of dielectric properties via the Clausius-Mossotti local field [67, 68] conditions. Generally, this model is limited to non-dipole liquids or hypothetical dielectric systems with (almost) non-interacting permanent dipole moments [69-72]. For such a case, the so-called Mosotti Catastrophe [70, 71]:

$$\chi(T) = \varepsilon(T) - 1 = \frac{M}{T - T_C} \qquad \text{for } T > T_C \tag{10}$$

where $\chi$ means dielectric susceptibility, directly reflecting polarization changes, and $T_C$ is the critical temperature.

This is a paradox that cannot occur in normal liquid dielectrics [69-72]. However, in the premelting domain there may conditions for its occurrence can appear, as indicated above. Natably, that Clausius-Mossotti local field obeys in ferroelectric systems, and in this case Eq. (10) is known as Curie-Weiss equation [73-76]. As recently shown [75], the validation of Clausius-Mossotti local field is associated with the inherent uniaxiality of basic ferroelectric systems [76]. Notable that under such conditions $\chi(T)$ is the susceptibility coupled to the order parameter (polarizability) [73-75], and Eq. (10) coincides with Lipovsky Eq. (9) [66].

For linseed, one can expect both factors, namely: (i) weakening of dipole-dipole interactions due to nano-constraints, and (ii) uniaxiality, which can indicate the pretransitional effect in the liquid phase (see Eq. 5).

For linseed oil, a unique post-freezing effect also exists. In the opinion of the authors, it can be associated with the (very) weakly discontinuous character of the liquid-solid phase transition. For such a transition, one can expect pretransitional fluctuation on both sides of the phase transition. However, in the solid phase, they are dumped by the solidification. Notwithstanding, one can expect that when cooling below $T_f = T_m^{cool}$ the solidification/crystallization is shaped by fluctuations, which leads to its formation as grains covered by liquid nano-layers.

## 2b. The low-frequency domain: supercriticality and the solid state premelting and post-freezing effects

The value of dielectric permittivity strongly increases for lowering frequency below the static domain. For $\varepsilon''(f, T)$ it can be discussed via the electric conductivity (see the Appendix). For $\varepsilon'(f, T)$ the problem remains puzzling, and often, such tests are discussed only heuristically [21, 77]. Figures 5 and 6 show dielectric constant changes in the surrounding of liquid–solid transition, for frequencies lowered to even to $f = 1 Hz$. In the liquid phase, evolutions of $\varepsilon(T)$ follow the pattern described by Eq. (5), associated with the crossover $d\varepsilon(T)/dT > 0 \leftarrow d\varepsilon(T)/dT < 0$. The range of such behavior decreases with lowering the measurement frequency. It can be associated with the loss of the ability to register the presence of supercritical fluctuations, related to the increase in the time scale described by Eq. (1a), and the rise of the observation time scale in LF domain. The latter is proportional to the inverse of the measurement frequency.

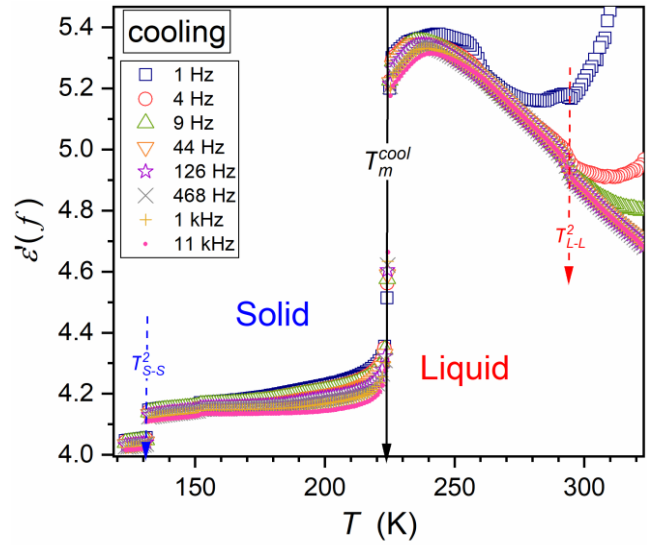

Figure 4    Temperature changes of the real part of dielectric permittivity in linseed oil on cooling from the liquid phase for the set of frequencies: from the static domain ($f = 11 kHz, 1 kHz$) to the low-frequency (LF) domain. Arrows indicate directly manifesting phase transitions.

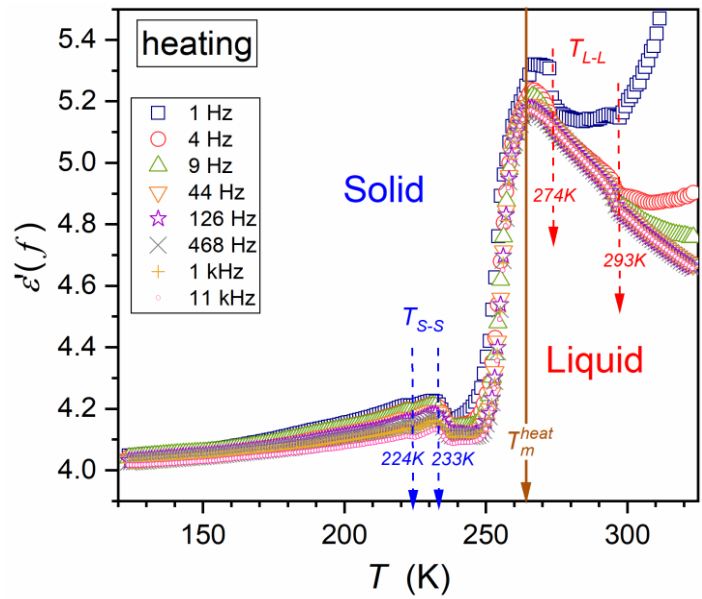

Figure 5    Temperature changes of the real part of dielectric permittivity in linseed oil on heating from the solid phase for the set of frequencies: from the static domain ($f = 10 kHz, 1 kHz$) to the low-frequency (LF) domain. Arrows indicate directly manifesting phase transitions.

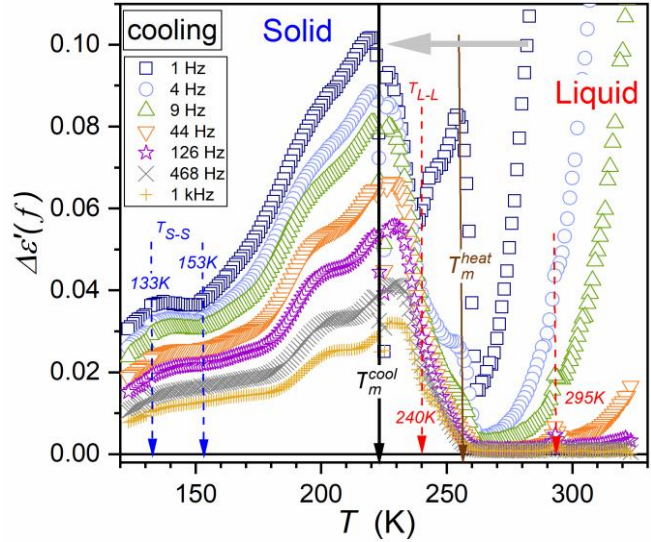

Figure 6    Linseed oil: temperature changes of the low-frequency contribution of the real part of dielectric permittivity $\Delta\varepsilon'(f) = \varepsilon'(f) - \varepsilon'(10kHz)$, obtained by subtracting the reference static value for $f = 10kHz$. Arrows indicate visible phase transitions.

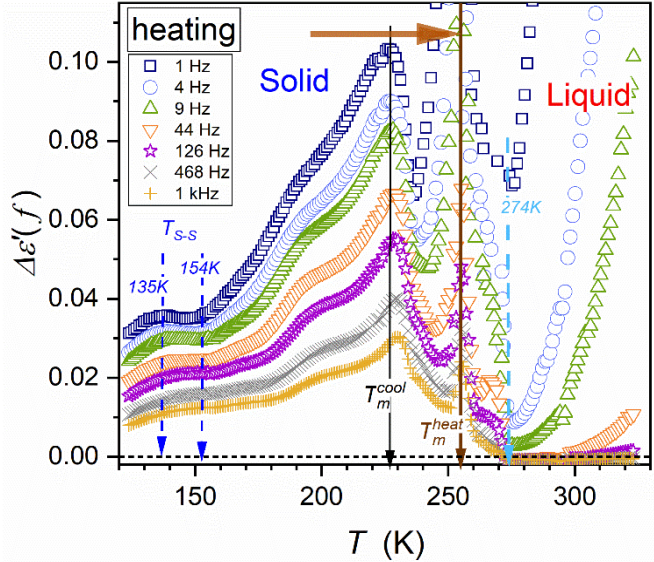

Figure 7    Linseed oil: temperature changes of the low-frequency contribution of the real part of dielectric permittivity $\Delta\varepsilon'(f) = \varepsilon'(f) - \varepsilon'(10kHz)$, obtained by subtracting the reference static value for $f = 10kHz$. Results for heating from the solid state. Apart from the 'real' melting temperature from the solid to the liquid state (black arrow), the location of the melting (freezing) temperature on obtained on cooling is indicated (light gray arrow). Arrows indicate visible phase transitions.

An additional factor that appears when lowering the measurement frequency is the increase of translational processes imacts, negligible in the static frequency domain. This factor can be taken into account by considering the following property [77]:

$$\Delta\varepsilon'(f = const, T) = \varepsilon'(f = const, T) - \varepsilon'(f = 10kHz, T) \qquad (10)$$

where $\varepsilon'(f = 10kHz) = \varepsilon$ is for the reference dielectric constant, with negligible impact of translation-related contributions. Such results are presented in Figures 6 and 7, on cooling and heating, respectively. This contribution (Eq. (10)) shows pretransitional impacts of translational processes on approaching freezing temperature or melting temperature. For the latter, it also shows strong quasi-pretransitional changes when passing $T_m^{cool}$ ($T_f$), despite the fact that no solid-liquid transition takes place for this path at $T_m^{cool}$. A similar phenomenon is visible when passing $T_m^{heat}$ in the liquid-phase cooling

### 3c.    Unique glassy dynamics

Both the real and imaginary parts of dielectric permittivity contais messages regarding the dynamics of processes, which can be registered by the BDS method. However, the analysis exploring $\varepsilon''(f)$ is more convenient in practice [15] (see also section 2 and Appendix). The analysis of the loss curves for subsequent temperatures enabled the determination of primary relaxation times. For linseed oil, it was possible to determine 3 such processes, and these results are presented in Figure 8, using the Arrhenius scale, the standard for showing such data [15].

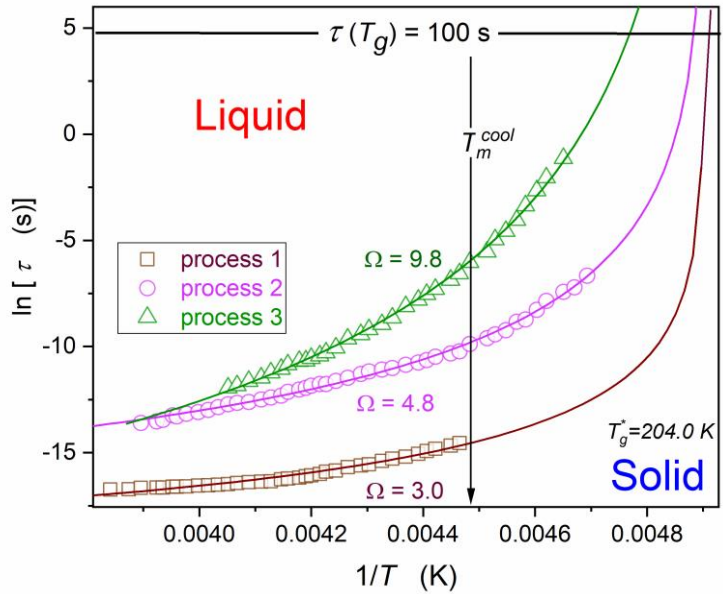

Figure 8  **Temperature changes of basic relaxation processes detected for the primary loss curve for the tested linseed oil.** Portraying curves are related to ADR Eq. (12), with parameters given in Table III, supported by the derivative analysis presented in Fig. 9.

Such evolutions are most often portrayed/analyzed using the Vogel-Fulcher-Tamman (VFT) super-Arrhenius equation [15, 16]. In refs. [16] it was proved that it is only an effective parameterization for most systems, not fundamentally justified. It is particularly true for systems with elements of the local uniaxial symmetry, and it seems to be the case of linseed, as shown in the analysis of dielectric constant pretransitional behavior above. In refs. [16, 78], the new route for analyzing such data, minimizing the application of the nonlinear fitting and testing the preferable way of their portrayal, was developed. First, experimental data are transformed to the form yield the apparent activation enthalpy $H_a(T)$ or equivalently, so-called apparent fragility (steepness index) $m(T)$ [78]:

$$m(T) = (T_g \ln 10) H_a(T) = c H_a(T) = \frac{H}{T - T_C^+} \Rightarrow$$

$$\Rightarrow 1/m(T), \ 1/H_a(T) \propto aT - b \tag{11}$$

where $T > T_g$, the amplitude $H = const$, and coefficients $a, b, c = const$; $H_a(T) = d\ln\tau(T)/d(1/T) = (d\tau(T)/\tau(T))/d(1/T)$; the singular temperature can be easily determined from the condition $\Rightarrow 1/m(T^+), \ 1/H_a(T^+) = 0$.

Eq. (11) was first noted for the supercooled, previtreous state of glass-forming liquids testing the evolution of the apparent fragility defined as $m(T) = d\log_{10}\tau(T)/d(T_g/T)$, where $T_g$ is for the glass temperature defined as $\tau(T_g) = 100s$ [78]. The universality of Eq. (11) was proved for 18 various systems so far [16, 78]. Notable, that the apparent fragility is directly proportional to the apparent enthalpy ( see Eq. (11); hence, its validity can be expected for any system with the super-Arrhenius (i.e., with the temperature-dependent activation energy) system. Figure 9 shows the mentioned universal behavior for three (primary) relaxation processes detected in linseed oil, with the same 'critical', singular temperature $T^+ \approx 204 \ K$.

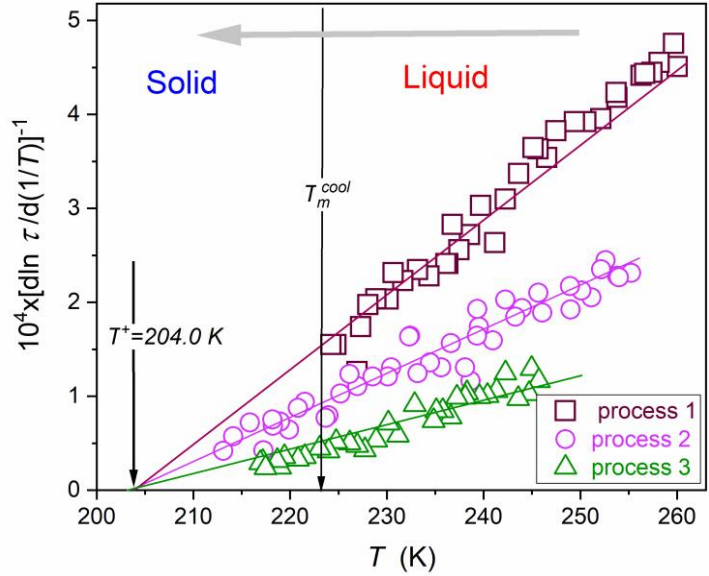

Figure 9  **Temperature evolutions of, the apparent activation enthalpy $H_a(T)$ reciprocals for primary relaxation processes detected in linseed oil.** Solid lines are related to the universal behavior given by Eq. (11).

Linking the empirical finding of Eq. (11) and definitions of the apparent enthalpy or the apparent fragility, the following relation was derived by one of the authors (ADR) [78]:

$$\tau(T) = C_\Omega \left(\frac{T - T^+}{T}\right)^{-\Omega} \left[\exp\left(\frac{T - T^+}{T}\right)\right]^\Omega = C_\Omega (t^{-1} \exp t)^\Omega \Rightarrow$$

$$\Rightarrow \ln(T) = \ln C_\Omega - \Omega \ln t + \Omega t \tag{12}$$

where $t = (T - T^+)/T = 1 - T_g^*/T$,

The above relation contains only 3 adjustable parameters, i.e., the same 'minimal' number as the popular VFT equation. However, the reliable value of $T^+$ can be estimated from the preliminary apparent activation enthalpy or apparent fragility analysis (see Fig. 9). As shown in Ref. [78] also, the exponent $\Omega$ and the prefactor $C_\Omega$ can be estimated from such a plot, reducing the final fitting of $\tau(T)$ experimental data to subtle adjustments. The unique form of Eq. (12) is worth indicating: it contains the critical-like and the 'activated' (i.e., exponential) terms. Solid curves in Figure 8 show the portrayal via Eq. (12), with parameters collected in Table III and partially recalled in the Figure. Generally, values of the exponent are located in the range $4 < \Omega < 28$, where the lower limit is typical for systems with the dominant local uniaxial symmetry [16, 78]. It is notable that two of three detected relaxation times can penetrate the solid phase, where they gradually diminish. A quation arises of it is associated with the decay of the post-freezing effect?

**Table III** Values of parameters in ADR Eq. (12), describing the basic evolution of relaxation times in linseed oil, shown graphically in Fig. 5.

| Parameter | $lnC_\Omega$ | $\Omega$ | $T_g^*$ (K) | $T_g$ (K) | $\Delta T_g^*$ (K) |
|---|---|---|---|---|---|
| Process 1 | -31.1 | 9.85 | 204.0 | 204.5 | 0.5 |
| Process 2 | -22.0 | 4.8 | 204.0 | 206 | 2 |
| Process 3 | -22.2 | 3.0 | 204.0 | 210 | 6 |

Figure 10 shows the temperature evolution of the DC electric conductivity, explicitly following the basic Arrhenius behavior, namely [15]:

$$\sigma^{-1}(T) = \sigma_\infty^{-1} exp\left(\frac{E_\sigma}{RT}\right) \tag{12}$$

with the same constant activation energy $E_\sigma$ on cooling and heating.

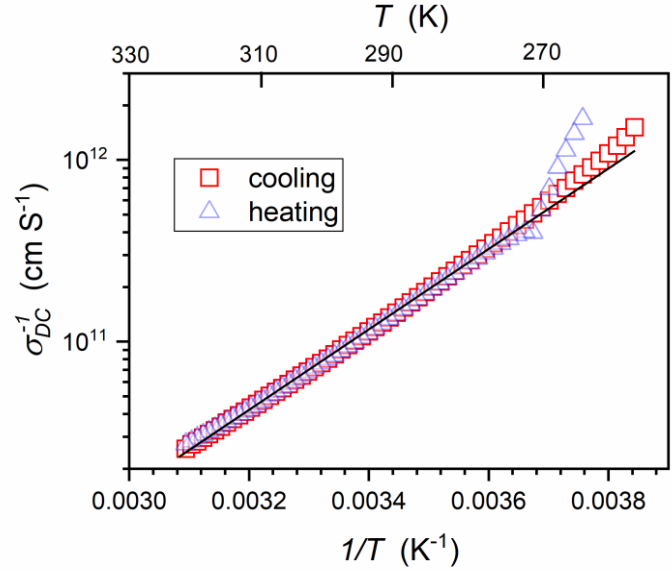

Figure 10 **Temperature evolution of the reciprocal of DC electric conductivity in linseed oil on cooling (*in blue*) and heating (*in pink*).** The straight line is related to the basic Arrhenius behavior. On passing $T \approx 275K$ the crossover to the SA-type behavior appears.

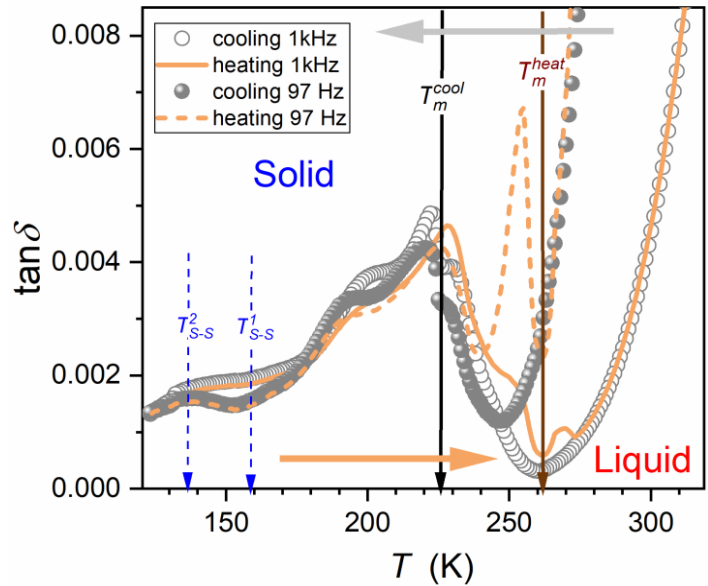

Figure 11 Temperature evolution of the loss factor $D = tan\delta$ in linseed oil on cooling from the liquid phase and on heating from the solid phase. Note explicit manifestations of $T_m^{cool}$ (i.e., $T_f$ ) and $T_m^{cool}$ (i.e., $T_m$), even if for the given path it is only passed without a phase transition.

The completed data set in the Appendix shows that for the direct surrounding or melting and freezing phase transitions, the unique shift of the DC conductivity domain to very low frequencies takes place. The real and imaginary components of dielectric permittivity can be linked defining the loss factor, dierectly describing process energy, which can be associated with heating, for instance [69, 72, 79, 80]:

$$D = tan\delta = \frac{i_{loss}}{i_{loss}+I} = \frac{\omega\varepsilon''+\sigma}{\omega\varepsilon'} = \frac{1}{Q} \qquad (13)$$

where $I$ is for the total current through the sample and $i_{loss}$ is related to its loss and the quality factor Q is the the ratio of the initial energy stored to the energy **lost** in one cycle for the given frequency $f$.

Consequently:

$$\varepsilon^* = \varepsilon' - i\varepsilon'' = \varepsilon'(1 - i \times tan\delta) \qquad (14)$$

This magnitude enables the estimation of the power loss $P$, which can be converted to the heat:

$$P = Qtan\delta = \omega CV^2 tan\delta = \varepsilon_0\varepsilon''E^2 \qquad (15)$$

where $\omega = 2\pi f$

## 4. Conclusions

The initial motivation for carrying out the research presented in this work was to determine the precise, and so far missing, characterization of the dielectric properties of linseed oil in a broad temperature range. However, studies revealed very strong and long-range supercritical changes in dielectric constant, resembling ones observed in the isotropic liquid phase of rod-like liquid crystalline compounds. Its influence reaches even room temperature and can be significantly greater and reach even higher temperatures, thanks to admixtures that can significantly change the phase transition temperature. This is also one of the ways, apart from changing the temperature, to control the level of the discovered supercriticality in linseed oil. In current applications based on the vicinity of the liquid-gas critical point (mainly in $CO_2$), the area associated with this phenomenon is the basis of supercritical technologies briefly characterized in the Introduction. The unusual features related to the supercritical region and its occurrence in linseed oil raise the question of whether the unique health-promoting properties of linseed oil are related only to its components, as previously believed, and not to the influence of supercriticality? This factor confirms the openness of innovative 'materials engineering' to enhance and perhaps even induce targeted effects of the health-promoting effects of linseed oil. In the case of the latter, it may be important, for example, to precisely determine the characteristics of the supercritical evolution of the dielectric constant (Eq.5), which can be combined with the Kirkwood relation (Eq, (2)). It is also possible to complete the extended Noyes-Whitney relation (Eq. (4)) by determining the evolution of DC electric conductivity (Eq. 12). The possibilities mentioned here related to finding and characterizing supercritical changes in the dielectric constant in linseed oil may also have broader significance in other applications of linseed oil, from the main component of far to 'ecological' transformer oil. There is also a potential for using linseed oil as a qualitatively new and ecological carrier supercritical agent in innovative supercritical extraction or processing technologies. For these applications, it may be important that the supercritical changes in the dielectric constant in linseed oil are qualitatively more significant than in the supercritical $CO_2$ gas phase. Moreover, processing occurs in the liquid phase of eco-friendly natural material.

The uniqueness of the supercritical effect in linseed oil mentioned here lies in its connection with the liquid-solid discontinuous transition, which is a significant novelty. In general, melting/freezing discontinuous transition is a common and extremely important phenomenon in nature and technological applications. However, unlike continuous phase transitions, it has remained a great cognitive challenge for over 100 years. Long-range pretransitional effects and their effective parameterization became the inspiration and validation for Critical Phenomena Physics, one of the largest universal studies in which the parameterization of experimental results for long-range phase transitions occurs. The problem with melting/freezing discontinuous transition is the practical lack of such pretransitional effects. Only weak premelting effects on heating from the solid state are detected. However, their parameterization seemed to be impossible.

Nevertheless, the premelting effect has become the main inspiration for theoretical models. A special role is played by the model of crystalline grains covered with nano-layers of a quasi-liquid, which leads to final fragmentation and melting. However, the lack of check-in relations for experimental validation has become a fundamental problem for the development of theoretical models. The exception was the Lipovsky model mentioned above. However, his predictions were for nano-layers of liquid, which were only a tiny part of melting solid, so validating the model seemed difficult, if at all possible.

This report shows that linseed oil has a premelting effect for dielectric constant and a significantly more significant post-freezing effect on cooling. Its parameterization is consistent with the Lipovsky model, as shown by the discussion indicating the appearance of quasi-negative pressure in nano-layer between solid grains and the occurrence of Mossotti Catastrophe type behavior. According to the authors, the unique appearance of the post-freezing effect in linseed oil may be related to the slightly discontinuous nature of the phase transition, for which 'critical' fluctuations should be expected to appear on both sides of the phase transition. Below $T_f$ this can lead to the 'granulated solidification', with liquid nano-layers' surroundings of solid-state granules. Interestingly, there are hallmarks of melting temperature in the way towards $T_f$ in the liquid phase and for the freezing temperature in the way to the melting temperature for the solid phase. These effects are especially visible for $D = tan\delta$, a quantity directly related to possible energy processes in linseed oil. The results show a surprising relationship of $T_f$ with the maximum of $D$ and $T_m$ with the minimum $D$, with a significant frequency dependence. These changes are visible far from the mentioned characteristic temperatures. This report shows 3 primary relaxation processes in the liquid phase, with 2 capable of limited penetration into the solid phase. The apparent activation enthalpy associated with these relaxation times, essentially a measure of the relative changes in the relaxation time in subsequent temperature intervals, shows universal changes related to the same quasi-critical singular temperature $T_C \approx 204 K$. The facts cited in this summary indicate the importance of linseed oil as a model system of fundamental importance. It was also important to conduct research using BDS because, with this method, the response from the liquid dielectric may be qualitatively more significant than that from the solid. For premelting and post-freezing effects in the solid state, this eliminates the problem of the tiny volume occupied by liquid nano-layers between solid-state grains.

**Acknowledgments**

The presented research was supported by National Center for Science (NCN, Poland) by the NCN OPUS (Poland) grant, ref. 2022/45/B/ST5/04005, guided by S. J. Rzoska.

**Authors contribution:**

ADR proposed the concept, participated in the analysis, and paper writing, and supplemented Figures by data analysis results; SJR proposed the concept, participated in the analysis and paper writing; JL carried out BDS studies and prepared Figures.

**Institutional Review Board Statement:** Not applicable.

**Informed Consent Statement:** Not applicable.

**Data Availability Statement:** Data available on reasonable requests to the corresponding author.

**Conflicts of Interest:** The authors declare no conflicts of interest with respect to the research, authorship, and/or publication of this article.

**Appendix**

Below are the results of complex dielectric permittivity measurements on cooling in linseed oil presented (Fig. A1), to illustrate experimental data. Basic features of the spectrum are also indicated. In Figure A3 the same data are presented in electric conductivity presentation to show the appearance of DC electric conductivity (horizontal lines) and distortions from such behavior.

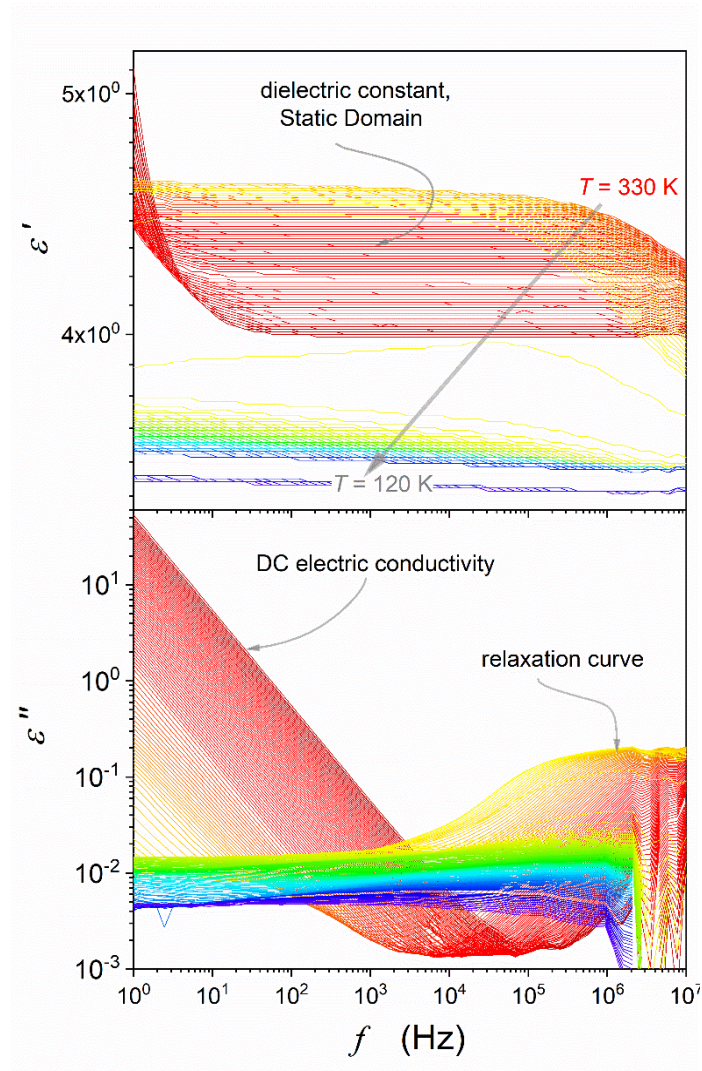

Fig. A1 **Real and imaginary components of dielectric permittivity presented for the tested range of temperatures in linseed oil.** The temperature shift is indicated by changing colors, from yellow (330 K), via red and green to deep-blue (120 K). Basic features of the spectrum are indicated. The horizontal part of $\varepsilon'(f)$ is related to dielectric constant and the sloped, low-frequency part of $\varepsilon''(f)$ to DC electric conductivity.

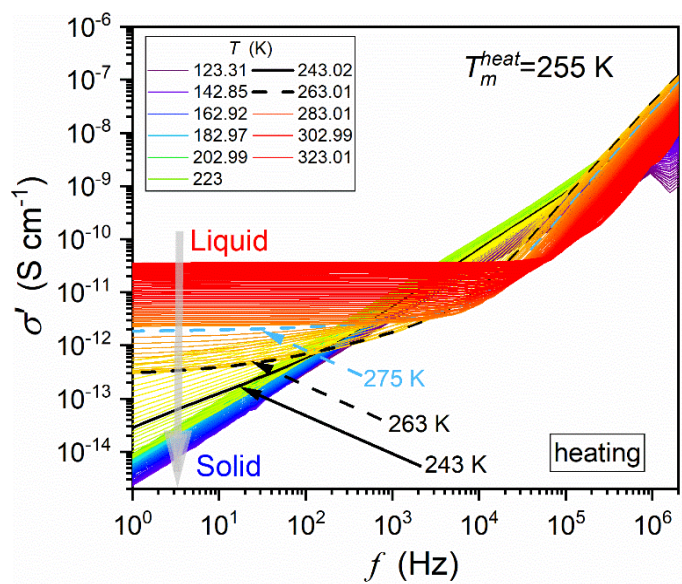

Figure A2  The real part of the complex electric conductivity obtained from $\varepsilon''(f)$ experimental data are given in Fig. A1. The horizontal behavior determines DC electric conductivity. The violation of such behavior is visible near melting and freezing temperatures. Results on heating from the solid phase.

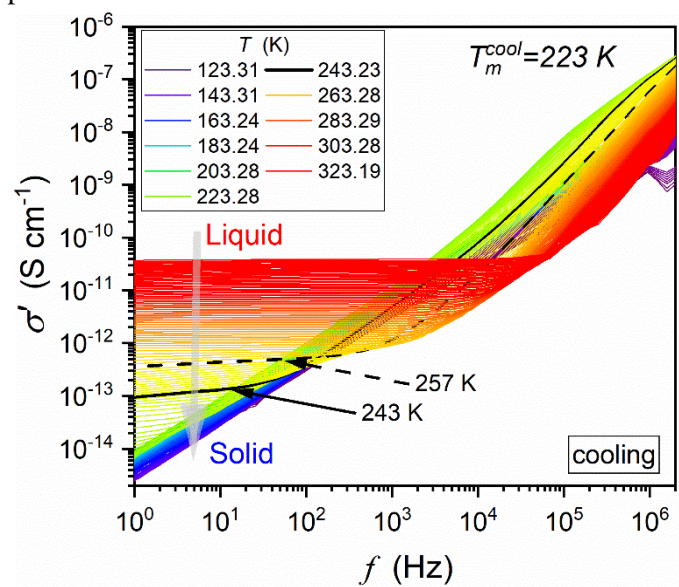

Figure A3  The real part of the complex electric conductivity obtained from $\varepsilon''(f)$ experimental data are given in Fig. A1 (see section 1). The horizontal behavior determines DC electric conductivity. The violation of such behavior is visible near melting and freezing temperatures. Results on cooling from the liquid phase.